\documentclass{moriond}

\usepackage{multirow}
\usepackage{amsmath,amssymb}

\def\be{\begin{equation}}
\def\ee{\end{equation}}
\def\bea{\begin{eqnarray}}
\def\eea{\end{eqnarray}}

\newcommand{\Photo}{}

\begin{document}
\vspace*{4cm}
\title{
UNSTABLE NEUTRINOS CAN RELAX COSMOLOGICAL MASS BOUNDS}

\author{ S. SANDNER }

\address{Instituto de F\'{\i}sica Corpuscular, Universidad de Valencia and CSIC, 
 Edificio Institutos Investigaci\'on, Catedr\'atico Jos\'e Beltr\'an 2, 46980 Spain}

\maketitle\abstracts{
The light neutrino masses are at present most stringently constraint via cosmological probes. In particular the Planck collaboration reports $ \sum m_\nu \leq 0.12\,\mathrm{eV}$ at $95\%$ CL within the standard cosmological model.
This is more than one order of magnitude stronger than the one arising from laboratory searches. The cosmological bound taken at face value excludes a plethora of neutrino flavour models which can successfully explain the neutrino oscillation data. The indirect nature of the cosmological bound, however, allows to relax the bound to up to $ \sum m_\nu \sim 1\,\mathrm{eV}$ if neutrinos decay on timescales shorter than the age of the Universe, $\tau_\nu \leq t_U$. 
We present how a decay of the type $\nu_i\to\nu_4\phi$ can be realized within general models of the minimal extended seesaw framework.
The idea is then explicitly realized within the context of a $U(1)_{\mu-\tau}$ flavour model.
}

\section{Introduction}
\label{sec:Introduction}

\begin{figure}[!ht]
\hspace{-0.0cm}  \includegraphics[width=1.0\textwidth]{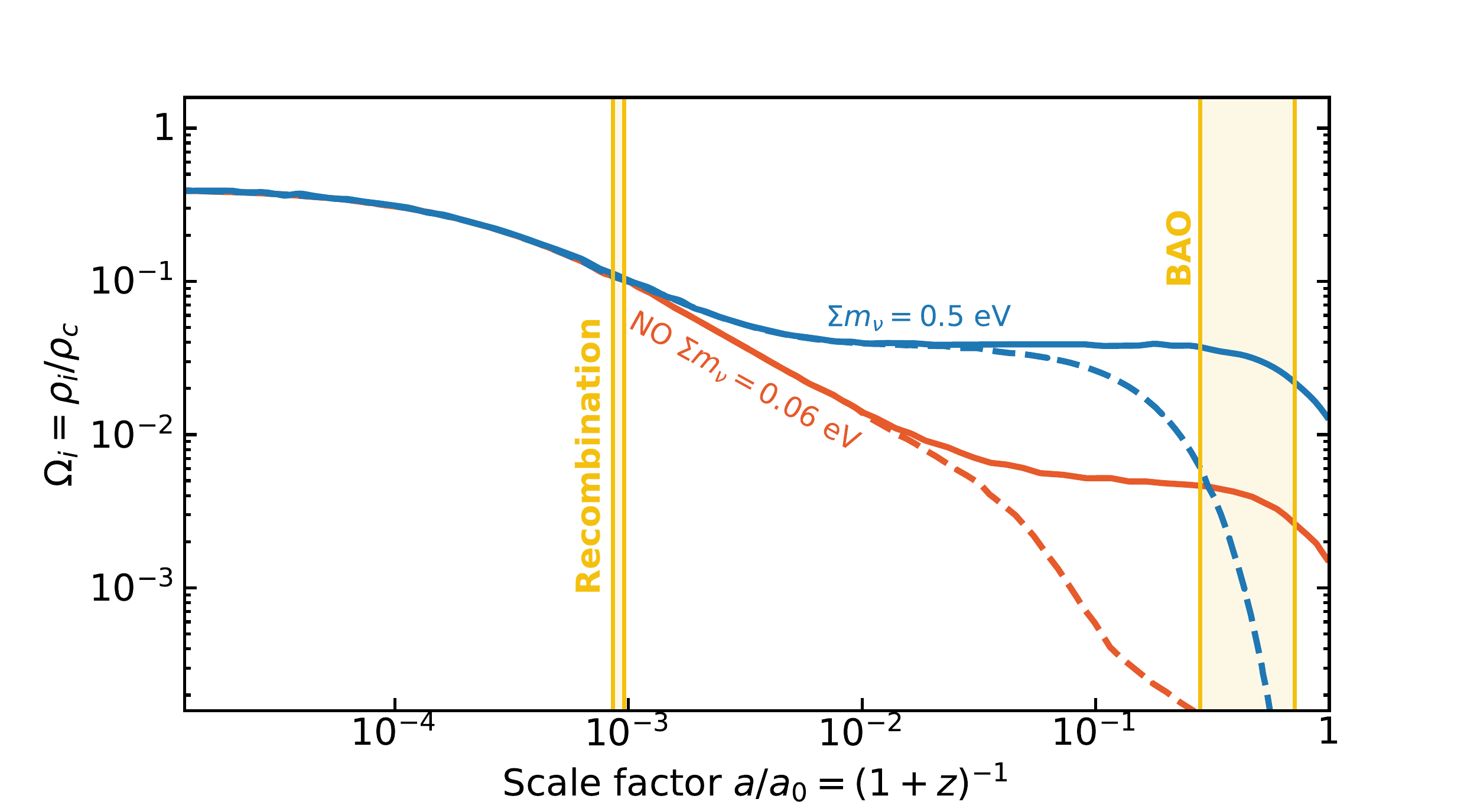}  
 \vspace{-0.0cm}
\caption{
Cosmological evolution of the normalized energy density for different neutrino masses. In dashed the corresponding evolution for cosmological fast decaying neutrinos is shown.
}\label{fig:omega_evolution}
\end{figure}

Neutrino oscillation experiments provided valuable information not only on the flavour mixing but also the neutrino mass square difference.
Due to the sensitivity to matter effects of the solar neutrinos we know that $\sqrt{\Delta m_{21}^2} \simeq 0.0086\,\mathrm{eV}$ but for the atmospheric neutrino mass splitting we only know $|\sqrt{\Delta m_{32}^2}| \simeq |\sqrt{\Delta m_{21}^2}| \simeq 0.05\,\mathrm{eV}$.
Although the neutrino hierarchy can be resolved within near future neutrino oscillation facilities, i.e. $m_{\nu_3} \leq m_{\nu_1}$ (inverted hierarchy (IH)) or $m_{\nu_1} \leq m_{\nu_3}$ (normal hierarchy (NH)), the neutrino absolute mass scale still remains undetermined and can not be addressed in this type of experiments.
The best laboratory constraint on the absolute neutrino mass scale comes from the KATRIN experiment that reports an upper bound of $\sum m_\nu \leq 2.7\,\mathrm{eV}$ at $95\%$ CL, but is expected to increase its sensitivity to $\sum m_\nu \leq 0.6 \,\mathrm{eV}$ in upcoming runs.
On the other hand, the cosmological neutrino mass bound is reported to be $\sum m_\nu \leq 0.12\,\mathrm{eV}$ at $95\%$ CL within the standard cosmological model $\Lambda CDM$. 
The main constraining power arises from the contribution of the neutrino to the hot dark matter content of our Universe.
In figure \ref{fig:omega_evolution} we summarize the evolutional history of the neutrinos over the course of the cosmological expansion.
Measurements requiere the neutrinos to be relativistic at recombination, making the evolutional history practically indistinguishable for 
$\sum m_\nu \leq 1.8\,\mathrm{eV}$ before recombination.
However, after recombination the evolution strongly depends on the absolute mass scale as this directly translates into a transition time between the relativistic and non-relativistic propagation. 
Non-relativistic neutrinos, since they behave as a matter component of our Universe, contribute to the hot dark matter content and are severely constraint by the baryonic acoustic oscillations.
Furthermore, the current cosmological bound is robust against common cosmological extension of the $\Lambda CDM$ model, as e.g. the inclusion of dark radiation or non-standard dark energy dynamics.
Nevertheless, a modification of the particle physics properties of the neutrino, as e.g. cosmological fast decays, can dilute the late time neutrino energy density and hence evade the main constraining power, which is represented by the dashed lines~\cite{Escudero:2020ped}.

\section{Categorizing Neutrino Decays}
\label{sec:category}

Neutrinos can decay already within the Standard Model (SM) of particle physics, which was first investigated in Ref.~\cite{Neutrino_Decay_Original}.
However, it was found that the decay rate exceeds by many orders of magnitude the current age of our Universe, i.e. $\tau_\nu^{\mathrm{SM}} \geq (G_{F}^2 m_\nu^5)^{-1} \gg t_U$.
First investigation of a possible neutrino decay via a Beyond the Standard Model (BSM) channel was done in Ref.~\cite{Invisible_Neutrino_Decay_Original} to explain the at the time present solar neutrino problem, which then, however, was successfully explained by corrections to the neutrino vacuum oscillation probability induced by matter effects.
The increasing experimental sensitivity and especially the in future expected reach motivates a detailed study on the field of neutrino decays to identify the required particle content, coupling strength and mass scale of the involved particles needed to result in $\tau_\nu \leq t_U$.

\begin{figure}[!ht]
\centering
\begin{tabular}{cc}
\hspace{-0.5cm} 
\includegraphics[width=0.45\textwidth]{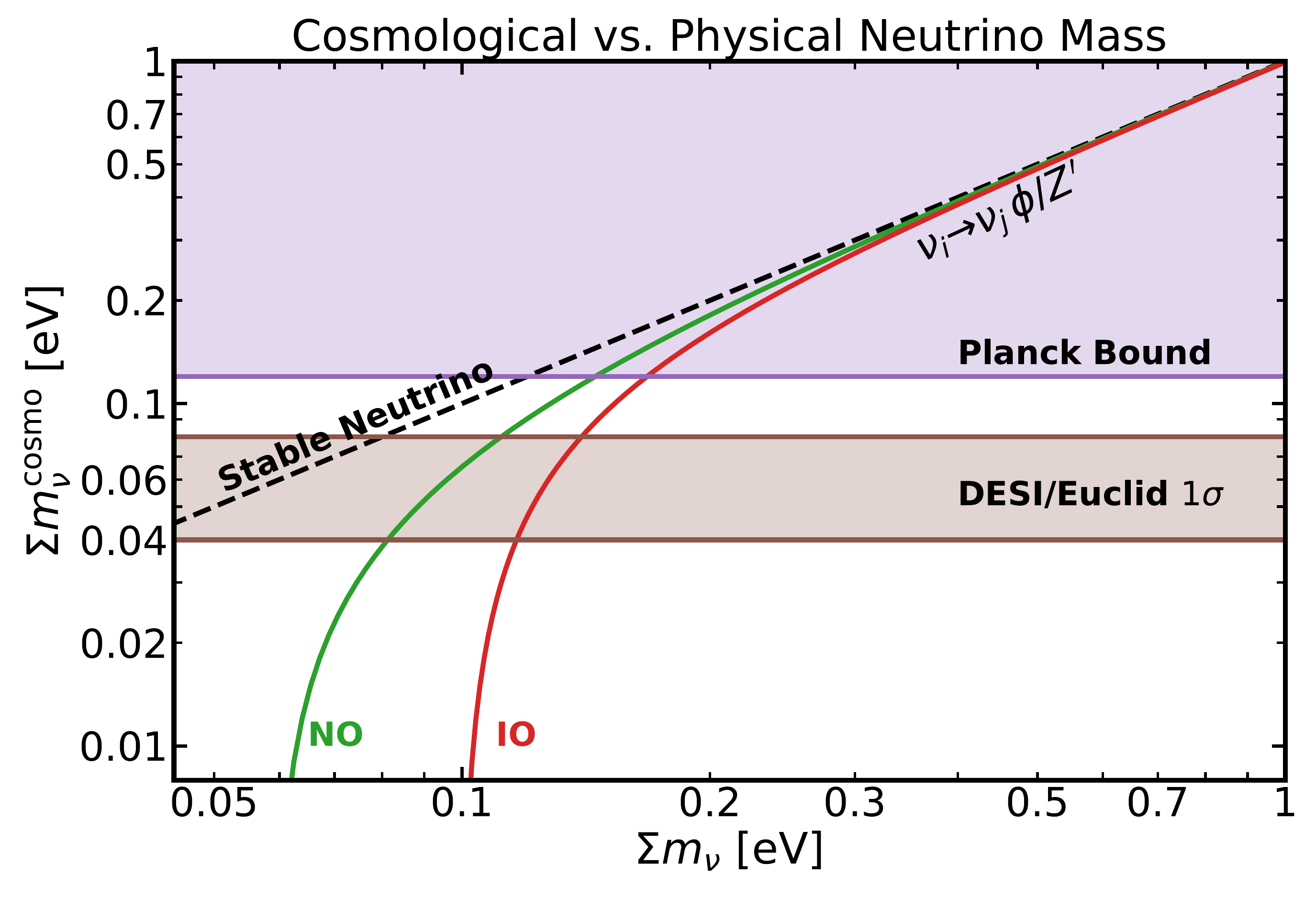} & 
\hspace{-0.5cm}  \includegraphics[width=0.45\textwidth]{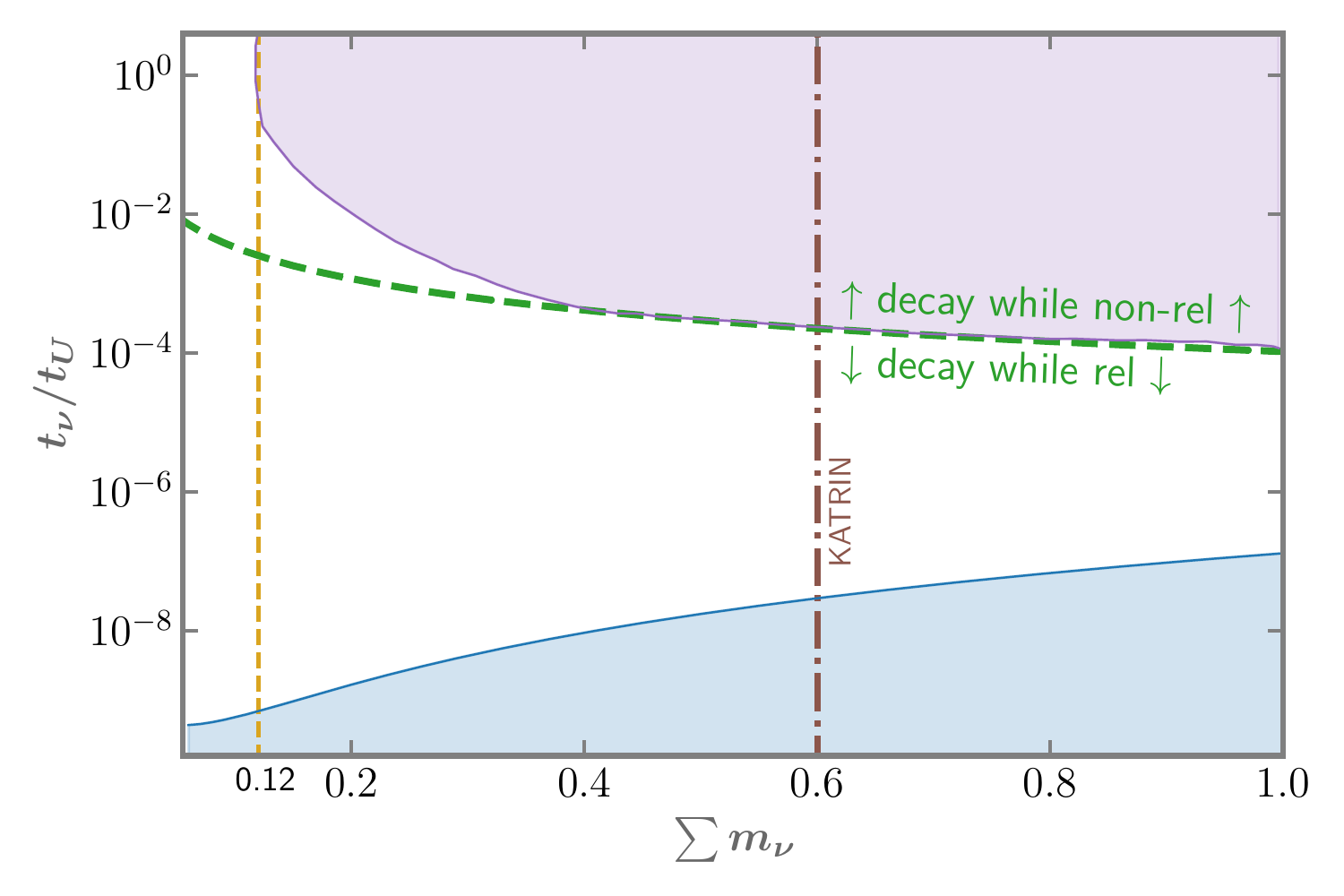}    \\
\end{tabular}
\vspace{-0.4cm}
\caption{ 
\textit{Left}: Relationship between the physical active neutrino mass and the one inferred from cosmological measurements. The correspondence is $1:1$ in the case neutrinos are cosmological stable. If $\tau_\nu \leq t_U/10$ the cosmological mass is given as a result of energy and angular momentum conservation by $\sum_\mu^{\mathrm{cosmo}} = 3 m_{\mathrm{lightest}}$.
\textit{Right}: Constraints on invisible neutrino decays for massless final state particles (dark radiation), see main text.
In yellow dashed we represent the current Planck bound on the neutrino mass.
}
\label{fig:mcosmo_mphys_small}
\end{figure}

Let us take a first look into the simplest BSM realization of (fast) neutrino decays, which is of the form $\nu_i \to \nu_j \phi/Z'$.
Here and in the following we will denote with $\nu_i,\nu_j$ active neutrino mass eigenstates and with $\nu_4$ the newly introduced sterile neutrino. 
Angular momentum conservation requieres the end product of the active neutrino decay to contain a scalar ($\phi$) or vector ($Z'$) boson.
Assuming that the exemplary decay is fast allows us to plot in figure~\ref{fig:mcosmo_mphys_small} the actual physical mass of the active neutrino against the one which would be inferred from cosmological measurements~\cite{Escudero:2020ped}. 
In the stable neutrino scenario the correspondence between both masses is one to one, represented by the black dashed line.
In contrary in the decaying neutrino scenario the physical neutrino mass is generically higher than the cosmological one and the actual relationship depends on the active neutrino hierarchy and is dictated purely by energy and angular momentum conservation.
In particular this can become important if future experiments as DESI \& EUCLID do not detect neutrino masses as this would contradict the classical normal and inverted hierarchy interpretation, whereas decaying neutrinos can bring both results and interpretation in accordance.

In full generality, we can categorize neutrino decays according to two criteria: i) the nature of the decay products and ii) the number of particles in the final state. 
This leads us to

\begin{enumerate}
	\item[\textit{i)}]  \textit{Nature of the decay products.}
\begin{enumerate}
	\item[\textit{a)}] At least another active neutrino mass eigenstate $\nu_j$.  \\
Since energy and angular momentum conservation ensure the lightest neutrino state to be stable, the cosmological constraint on $\sum m_\nu$ cannot be relaxed by more than $ 0.06\,\mathrm{eV}$ and $0.1\,\mathrm{eV}$ for NH and IH, respectively.
Although it can get very relevant in the future, as discussed before, we will not focus on this possibility, since the relaxation is tightly constraint.
	\item[\textit{a)}] Only BSM species.  \\
In this case, the constraint on $\sum m_\nu$ can be significantly relaxed to the level of $\sum m_\nu \leq 1\,\mathrm{eV}$ at 95\% CL, provided that the BSM particles are massless. 
This is frequently called neutrino decay into dark radiation.
The exact bound is subject to ongoing investigation and recently updated in Ref.~\cite{Neutrino_CMB_Bounds}.
New experimental data sets, including the latest Planck 2018+BAO+Pantheon, and a refined analysis significantly strengthened the bound on a possible cosmological neutrino decay while being non relativistic. 
The main constraining power arises from the time modification of the gravitational pull and the resulting modification of the Hubble parameter and hence a modified late time large scale structure formation.  
On the other hand, if $1/\tau_\nu \geq H_0 \sqrt{\Omega_m} (\sum m_\nu/(9 T_{\nu,0}))^{3/2}$, neutrinos decay while being relativistic and the same reasoning can not be applied.
However, decays in such a regime can be constrained by noticing that decay and inverse decay processes alter the neutrino free-streaming properties to lower scales which results in a phase shift of the high multipole moments in the cosmic microwave background power spectrum.
The resulting bound in the phase space of active neutrino lifetime and its mass is shown in figure~\ref{fig:mcosmo_mphys_small}.
\end{enumerate}

	\item[\textit{i)}]  \textit{Number of particles in the final state.}
\begin{enumerate}
\item[\textit{2)}] 2-body decays.\\
Angular momentum conservation requires the decay products to be one fermion and one boson with spin~$0$ (scalar $\phi$) or~$1$ (vector $Z'$). Regarding the fermion, we will consider two possibilities: a light neutrino mass eigenstate $\nu_i$ or a sterile neutrino $\nu_4$.

\item[\textit{3)}] 3-body decays and beyond.

For sufficiently massive bosons, i.e. $m_{\phi,\,Z'} > m_{\nu_i}$, any 2-body decay is kinematically closed. However, such a boson can mediate off-shell a 3-body decay, which becomes the dominant channel. 
In a model independent analysis only the decay of $\nu_i \to \nu_4\nu_4\nu_4$ allows to relax the cosmological neutrino mass bound in a broad region of parameter space~\cite{Escudero:2020ped}. 
However, in realistic neutrino models the mixing between active and sterile neutrino states will close the viable window completely. 
Hence, we will not focus on 3-body decays and beyond here.

\end{enumerate}
\end{enumerate}

Therefore, according to this classification and discussion therein, six distinct neutrino decay topologies are possible out of which only the decay $\nu_i \to \nu_4 \phi/Z'$ can relax the bound to up $\sum m_\nu \sim 1\,\mathrm{eV}$ in concrete model realizations.
This will be the channel of interest for the rest of the work.
The effective Lagrangian describing the new interaction with a scalar boson is
\begin{eqnarray}
\label{eq:Lag_scalar}
\mathcal{L}^{\phi} & \supset - \frac{\phi}{2} \, \left[  \overline{\nu_i} \left(h_{ij} + i \lambda_{ij} \gamma_5  \right) \nu_j +  \overline{\nu_i} \left(h_{i4} + i \lambda_{i4} \gamma_5  \right) \nu_4 + \overline{\nu_4} \left(h_{4i} + i \lambda_{4i} \gamma_5  \right) \nu_i \right] +  \mathrm{h.c.}\,,
\end{eqnarray}
and with a vector boson
\begin{eqnarray}\label{eq:Lag_vector}
\mathcal{L}^{Z'} & \supset  -\frac{Z'_\mu}{2} \left[ g_{ij}^{L}\, \bar{\nu}_i \gamma^\mu P_L  \nu_j+  g_{i4}^{L} \,\bar{\nu}_i  \gamma^\mu P_L \nu_4 + g_{4i}^{L} \,\bar{\nu}_4  \gamma^\mu P_L \nu_i \right]+ \mathrm{h.c.}\,,
\end{eqnarray}
where $P_{L(R)}$ is the left (right) handed projection operator and $g^{L(R)}$ represent left (right) handed couplings. 
Note that we neglect the direct coupling between the sterile neutrinos because it can only be phenomenological relevant for $3-$body decays (and beyond) which, however, are strongly disfavored as discussed above.
The decay rate for the scalar interaction is then given by
\be
\label{eq:gamma_i4_pheno}
 \Gamma \simeq \frac{\lambda_{i4}^2}{16\pi} \, m_{\nu_i} \simeq  t_{U}^{-1} \left(\frac{m_{\nu_i}}{0.3~\mathrm{eV}}\right) \left(\frac{\lambda_{i4}}{5\times 10^{-16}}\right)^2\,,
\ee
and for the vector interaction
\be
\Gamma \simeq \frac{g^{L}_{i4} {}^2}{16\pi} \, \frac{m_{\nu_i}^3}{m_{Z'}^2} \simeq t_{U}^{-1}\left(\frac{m_{\nu_i}}{0.3~\mathrm{eV}}\right)^3  \left( \frac{400 \,\mathrm{TeV}}{m_{Z'}/g_{ij}^L}\right)^2\,.
\ee
Note that the translation of the Lagrangians given in equations~(\ref{eq:Lag_scalar}) and (\ref{eq:Lag_vector}) into the neutrino flavour basis, active ($\nu_\alpha$) and sterile ($\nu_s$), can be realized via a rotation of the type $\nu_\alpha=U_{\alpha i}\,\nu_i+\theta_{\alpha 4}\,\nu_4$, $\nu_s=\theta_{s i}\,\nu_i+\theta_{s 4}\,\nu_4$.
Since the mixing between the sterile and active neutrinos $\theta_{\alpha 4}$ and $\theta_{s i}$ should be small, $U_{\alpha i}$ with $\alpha=e,\mu,\tau$ and $i=1,2,3$ is given by the PMNS matrix, up to subleading corrections driven by the active-sterile neutrino mixing.

On the level of the effective Lagrangian couplings can be taken to be unrelated but in UV complete theories couplings can be correlated via mixing patterns which may have non-trivial origins.
A first naive expectation is that one should expect either one of the two 
\begin{itemize}
\item $(\lambda_{ij}: \lambda_{i4} : \lambda_{44})  \simeq (1:\theta:\theta^2)$: Only active neutrinos interacting directly with a new force carrier and $\nu_4-\phi/Z'$ interaction arising via active-sterile neutrino mixing $\theta$. 
\item $(\lambda_{ij}: \lambda_{i4} : \lambda_{44})  \simeq (\theta^2:\theta:1)$: Sterile neutrinos interacting directly with a new force carrier while active neutrinos interact only via the mixing $\theta$.
\end{itemize}
In the following section we will construct a model based on a minimal extension of the generic seesaw neutrino mass framework which does not show the above mentioned hierarchical pattern but leads to viable $\nu_i \to \nu_4 \phi$ decay.

\section{Concrete Model Realization}

Successful neutrino models have to show a mass and flavour structure in accordance with the latest oscillation data~\cite{Esteban:2020cvm}.
Many of these models realize such a pattern via imposed flavour symmetries which introduce symmetry protected zero modes in the neutrino mass matrix.
A commonly considered two-flavour symmetry leads to two-zero textures which can successfully explain neutrino oscillation data but predict $\sum m_\nu \geq 0.12~\mathrm{eV}$ in $5$ out of $7$ viable realizations~\cite{Two_Zero_Texture}.
Furthermore, even though other models may not strictly predict neutrino masses in conflict with the current cosmological bound many of them accommodate for $\sum m_\nu \geq 0.12~\mathrm{eV}$ across large regions of the parameter space.
We will focus on the case of a neutrino model with an underlying $U(1)_{\mu-\tau}$ flavour symmetry and incorporate an invisible neutrino decay via a simple extension of the seesaw framework. 

\subsection{Neutrino Decay within the Seesaw Framework }

The realization of the decay $\nu_i \to \nu_4 \phi$ should not spoil the active neutrino mass generation mechanism by the introduction of the new BSM states.
In particular, a large mixing between the active neutrino and the BSM states should be prevented to be in accordance with current bounds on light sterile neutrinos. 
Hence, the task is to generate a (nearly) massless sterile neutrino.

This, for example, can be realized by adding an extra fermion singlet $S_L$ and one complex scalar singlet $\Phi$ to the usual seesaw model field content consisting of $\geq 2$ Majorana type right-handed neutrinos.
The stability of the proposed model then is guarantied by an additional global symmetry $U(1)_X$.
Assigning the new fields $S_L$ and $\Phi$ with opposite $U(1)_X$ charges while having the rest of the field content uncharged the unique symmetry allowed term is $y\Phi \bar{N}_R S_L$.
In the moment when the global $U(1)_X$ symmetry will by dynamically broken the complex scalar $\Phi$ obtains a vacuum expectation value $v_\Phi$, rendering a massless goldstone boson -- the Majoron -- and a massive real part of order $v_\Phi$.
Working in the flavour basis, the $7\times7$ neutrino mass matrix after symmetry breaking is given by
\begin{eqnarray}
\label{eq:Mnu}
 M_\nu=\begin{pmatrix}
0 & m_D & 0\\
m_D^t & M_R & \Lambda\\
0 & \Lambda^t &0
\end{pmatrix},
\end{eqnarray}
which is often referred to as the minimal extended seesaw.
Here $m_D$ and $M_R$ are the $3\times3$ Dirac and Majorana mass matrices, representing the usual seesaw model.
The new piece is the $3\times1$ matrix $\Lambda$ whose elements are given by $\Lambda_\alpha = y_\alpha v_\Phi$.
The eigenvalues of the matrix represent the mass eigenstates of the different particles at hand and the diagonalization under the assumption of a scale hierarchy of the form $\Lambda \ll m_D \ll M_R$ leads to 
\begin{eqnarray}
 m_i&\simeq& m_D^2/M_R,\;\;\;\nu_\alpha\simeq \nu_i+\left(m_D/M_R\right)N_i-\left(\Lambda/m_D\right)\nu_4\,,
\nonumber\\
M_i&\simeq& M_R,\;\;\;\;\;\;\;\;\;N_R^c\simeq -(m_D/M_R)\nu_i+N_i\,,
\nonumber\\
m_4&=& 0,\;\;\;\;\;\;\;\;\;\;\;\;\;\;S_L\simeq (\Lambda/m_D)\nu_i+(\Lambda/M_R)N_i+\nu_4\,.
\end{eqnarray}
A striking feature is the immediate identification that the active neutrino mass generation is dominantly achieved via the typical seesaw mechanism, while the sterile state at tree level stays exactly massless. 
Corrections to the active neutrino mass arising from the newly introduced sterile neutrino are heavily suppressed and of order $\mathcal{O}(\Lambda^2/M_R)$ as long as we are in the limit of the assumed scale hierarchy.
The flavour composition is also such that mixings between different states are always suppressed by $x/y$ where $x$ is some low-scale and $y$ some high-scale.
This allows the model to be basically unaffected by BBN and neutrino oscillation constraints.
The strongest constraint on the active-to-sterile mixing is derived from cosmological BBN data to be $ |\theta_{\alpha4}|^2  \lesssim 4\times 10^{-6} \mathrm{eV}/\sqrt{|\Delta m_{4i}^2|}$, while laboratory neutrino oscillation data constrain the mixing to be $ |\theta_{\alpha4}|^2  \lesssim \mathcal{O}(0.01)$.
Note that as a result of the assumed scale hierarchy we have $m_{\nu_4}^2 - m_{\nu_i}^2 < 0$, instead of the commonly considered scenario of $m_{\nu_4}^2 - m_{\nu_i}^2 > 0$ to explain neutrino oscillation anomalies.
Hence, in our model $\sqrt{|\Delta m_{4i}^2|} \simeq m_{\nu_i} \lesssim 1~\mathrm{eV}$ which leads to $ |\theta_{\alpha4}|^2 \sim \Lambda/m_D \lesssim 10^{-3}$, which is equivalent to requiring a minimal scale hierarchy for the model to work.

To connect to the discussion in the previous section we proceed by identifying the BSM neutrino interactions in the mass basis which are given by expanding the unique new term in the interaction Lagrangian
\begin{eqnarray}
 \Delta  \mathcal{L}&=& \frac{y}{2}\Phi\overline{N_R}S_L+h.c \supset -\frac{y}{2}\frac{\Lambda}{M_R}\overline{\nu_i^c}(\sigma-i\gamma_5\phi)\nu_j-\frac{y}{2}\frac{m_D}{M_R}\overline{\nu_i^c}(\sigma-i\gamma_5\phi)\nu_4
 \nonumber\\
 &+&\frac{y}{2}\overline{N_i}(\sigma-i\gamma_5\phi)\nu_4 -\frac{y}{2}\frac{\Lambda}{m_D}\overline{N_i}(\sigma-i\gamma_5\phi)\nu_j+h.c\,,
  \label{eq:DeltaL}
\end{eqnarray}
in which we denote with $\phi$ the Majoron and $\sigma$ the massive radial component of $\Phi$.
Now comparing to the phenomenological Lagrangian given in eq.~\eqref{eq:Lag_scalar} it is easy to identify the relation
\begin{align}\label{eq:mapping}
\lambda_{ij} \simeq  y \,\Lambda/M_R, \,\,\,\, 
\lambda_{i4}\simeq y\, m_D/M_R,\,\,\,\,\lambda_{44}\simeq0\,,
\end{align}
which does not match any naive expectation for the coupling hierarchy as presented in section~\ref{sec:category}.
The sterile states among each other are non-interacting, while active-active and active-sterile interactions are effective.
The dominant coupling is indeed the desired active-to-sterile portal which can be parametrized using the typical seesaw scales as
\begin{equation}
\label{eq:lambda_i4_MES}
\lambda_{i4}\simeq y\cdot 10^{-13}\sqrt{\left(\frac{m_{\nu}}{0.1\rm{eV}}\right)\left(\frac{10^{16}\rm{GeV}}{M_R}\right)}\,.
\end{equation}
Note that although the model additionally acomodates for three body decays if $m_\sigma \gtrsim m_\nu$ it can not be used to relax the cosmological neutrino mass bound because $\lambda_{44} \simeq 0$.
Having discussed the general strategy to account for neutrino decay within minimal extended seesaw model we will turn to a concrete model realization in the following.

\subsection{Flavour Symmetric Model}

An example for a neutrino model which satisfies all neutrino oscillation data but predicts $\sum m_\nu = 0.146~\text{eV} - 0.227~\text{eV} \,(0.121~\text{eV} - 2.7~\text{eV}) \geq 0.12~\text{eV}$ at 1-$\sigma$ (3-$\sigma$) is the one based on a global $U(1)_{\mu-\tau}$ flavour symmetry. 
These type of models are also trivially anomaly free within the SM.
We shall adapt this model in order to illustrate a concrete UV-completion of the above discussed neutrino decay scenario within the minimal extended seesaw framework.
To this end we first investigate the standard $U(1)_{\mu-\tau}$ scenario before we add the minimal extended particle content.

\textit{Minimal $U(1)_{\mu-\tau}$ model allowing for $\nu_i \to \nu_j \phi$}
~\\
The minimal $U(1)_{\mu-\tau}$ model is equipped with only one SM singlet scalar $\varphi$ and $3$ heavy sterile neutrinos $N_i$ with $U(1)_{\mu-\tau}$ charges $Q(\varphi)=+1$, $Q(N_e)= 0$ and $Q(N_\mu) = - Q(N_\tau) = +1$.
After the dynamical breaking of the global $U(1)_{\mu-\tau}$ symmetry the CP-odd odd component of $\varphi$ appears a massless Goldstone boson $\phi$~\footnote{Note that when gravity is included into the theory we may end up with $m_\phi > 0$ but due to its weakness we can safely expect $m_\phi \ll m_\nu$.}.
Hence, the model allows for decays of the type $\nu_i \to \nu_j \phi$.
The coupling $\lambda_{ij}$ controlling the decay probability is generated via mixing between the heavy sterile and active neutrinos through the interaction $\mathcal{L} \supset - \mathcal{Y}_{e\mu}\varphi \bar{N_{e}^{c}} N_\mu - \mathcal{Y}_{e\tau} \varphi^\dagger \bar{N_{e}^{c}}N_\tau$ where $\mathcal{Y}$ denote the Yukawa couplings.
To relate to the phenomenological coupling as expressed in eq.~\eqref{eq:Lag_scalar} we use the standard seesaw relation for the mixing $\theta_{N\nu} \simeq \sqrt{m_\nu / M_N}$ and assume $\mathcal{Y}_{e\mu} \simeq \mathcal{Y}_{e\tau}$ to end up with
\be
\label{eq:lambda_ij_mu_tau}
\lambda_{ij} \simeq  \mathcal{Y}_{e\mu} \theta_{N_e\nu_i} \theta_{N_\mu \nu_j} +  \mathcal{Y}_{e\tau}  \theta_{N_e\nu_i} \theta_{N_\mu \nu_j} \simeq \frac{\sqrt{m_{\nu_i} m_{\nu_j} }}{v_{\mu-\tau}} \simeq 10^{-11} \frac{m_{\nu_i}}{0.1\,\mathrm{eV}} \frac{10\,\mathrm{GeV}}{v_{\mu-\tau}}\,.
\ee
The requirement of cosmological fast neutrino decay leads to
\be
   \sqrt{ \frac{16\pi}{f(m_{\nu_i},m_{\nu_j})} t_U}  \lesssim \lambda_{ij} \lesssim  \sqrt{\frac{32\pi}{f(m_{\nu_i},m_{\nu_j})} 10^{-8} t_U \left( \frac{m_{\nu_i}}{0.1\,\text{eV}}\right)^3} \,,
\ee
where we defined the $2-$body decay phase space function $f(m_{\nu_i}, m_{\nu_j}) = (m_{\nu_i} - m_{\nu_j})^3 (m_{\nu_i} + m_{\nu_j})/m_{\nu_i}^3$.
CMB constraints on the neutrino free streaming property sets the upper limit on the coupling.
Given the expression of the coupling above in eq.~\eqref{eq:lambda_ij_mu_tau} this maps to $ 1~\text{GeV} \lesssim v_{\mu-\tau} \lesssim 10~\text{TeV}$.
Interestingly, the required scale is just around the electroweak scale.
Recall that the relaxation for decays of type active-to-active is bounded by the mass splitting between the active neutrino mass eigenstates and hence we need to extend the model to account for a relaxation of $\sum m_\nu \sim 1~\text{eV}$.

\begin{figure}[!ht]
\hspace{0.15 \textwidth}  \includegraphics[width=0.6\textwidth]{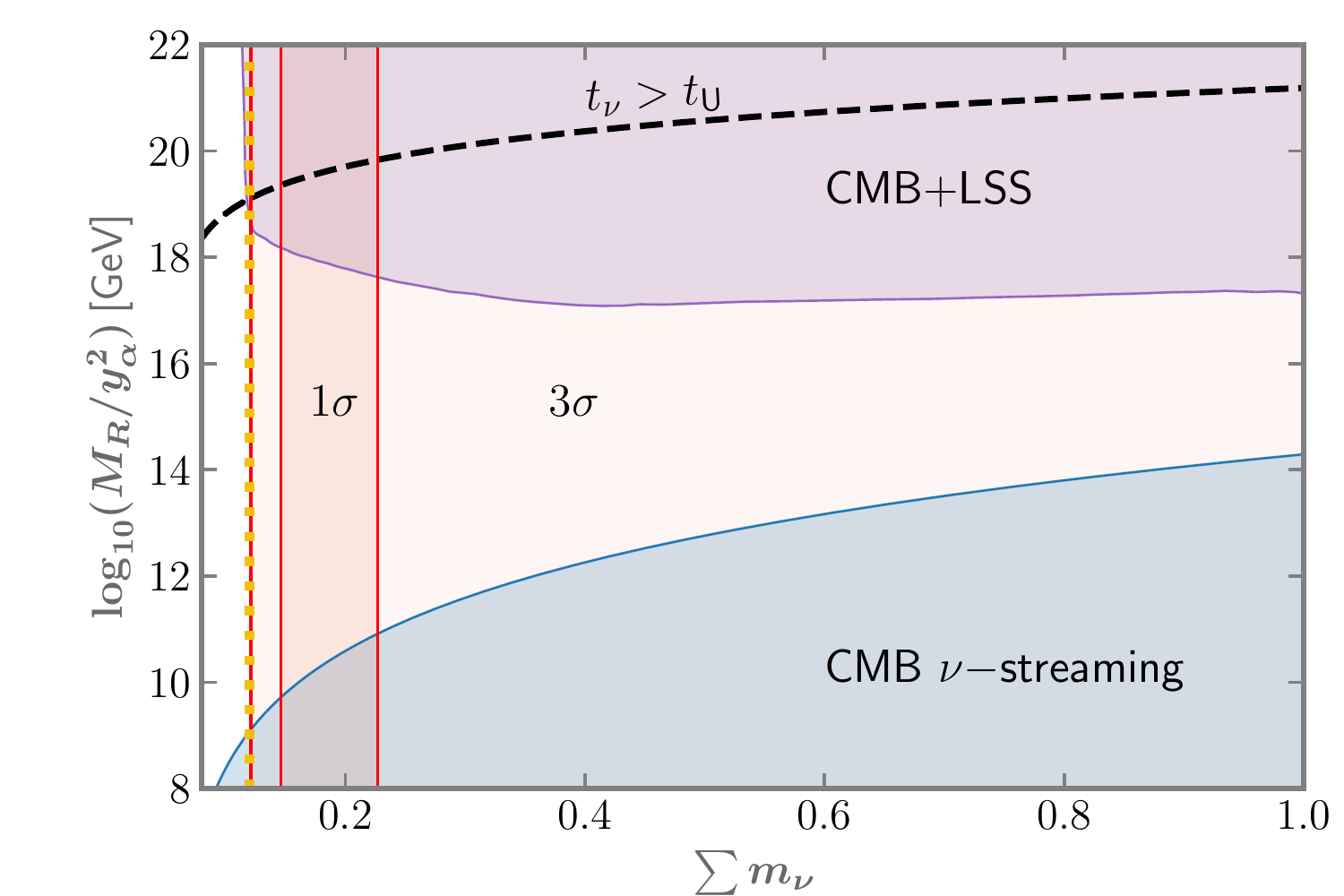}  
 \vspace{-0.0cm}
\caption{
Parameter space of the $U(1)_{\mu-\tau}$ model within the minimal extended seesaw framework. The red bands indicate the $1\sigma$ and $3\sigma$ prediction for neutrino masses within de model, generically being higher than the current Planck bound (yellow dashed). Neutrino decays within the model can relax the tension within various orders of magnitude in Yukawa coupling and Majorana mass scale.
}\label{fig:Model_Constraint}
\end{figure}

\textit{Minimal Extended $U(1)_{\mu-\tau}$ model allowing for $\nu_i \to \nu_4 \phi$}
~\\
Let us now turn to discuss an extension of the analyzed $U(1)_{\mu-\tau}$ within the minimal extended seesaw framework.
Following the general discussion around eq.~\eqref{eq:Mnu} we add one singlet fermion $S_L$ and one complex scalar $\Phi$ to the model.
Both are oppositely charged under some new global symmetry $U(1)_X$, but only $\Phi$ can carry a $U(1)_{\mu-\tau}$ charge in order to guarantee anomaly cancellations.
In order to allow for active-to-sterile neutrino conversion via the coupling $y\Phi\bar{N_R}S_L$ the $\Phi$ charge under $U(1)_{\mu-\tau}$ has to one of the three options $Q(\Phi) = 0,\pm1$, which leads to $\Lambda^T = \{\Lambda_e,0,0\}$ or $\Lambda^T = \{0,\Lambda_\mu, \Lambda_\tau\}$ respectively.
Combining the model independent result for the decay rate of $\nu_i \to \nu_4 \phi$, as given in eq.~\eqref{eq:gamma_i4_pheno}, with the constraint on the coupling arising from the minimal extended seesaw framework, as given in eq.~\eqref{eq:lambda_i4_MES}, we arrive at
\be
\Gamma_{\nu_i \to \nu_4 \phi} \simeq 10^{6} t_U^{-1} y^2 \left( \frac{\sum m_\nu}{1~\text{eV}} \right)^2 \left( \frac{10^{14}~\text{GeV}}{M_R} \right)\,.
\ee
We show in figure~\ref{fig:Model_Constraint} the physical parameter space in which the cosmological mass bound can be relaxed, together with the $1\sigma$ and $3\sigma$ prediction for the light neutrino mass within the considered model.
In particular, for Yukawa couplings of order $y\sim 1$ the neutrino mass bound can be relaxed for a Majorana scale of $10^{10}\,\text{GeV} \lesssim M_R \lesssim 10^{19}~\text{GeV}$, which points to a canonical realization of the seesaw mechanism. 
Thereby, potential cosmological constraints on the right handed Majorana neutrino can be evaded as well as its effects on low energy particle physics experiment would be suppressed in the same manner.

\section{Summary and Conclusion}

Neutrino masses are at present best constraint via the indirect inference from cosmological measurements, while direct laboratory bounds are currently one order of magnitude weaker.
However, the cosmological neutrino mass bound, although stable against typical modification of the background evolution, is strongly dependent on the underlying assumption of the particle physics nature of the neutrino.
The impact of cosmological fast neutrino decays onto the cosmic microwave background has been investigated in Ref.~\cite{Neutrino_CMB_Bounds}.
This allows to constraint the underlying particle physics parameters as i.e. the involved couplings and its minimal beyond the Standard Model realization.
General constraints can be derived via a model independent analysis which can be used to map to any concrete UV completed model realization.
In particular, flavour models of the type $U(1)_{\mu-\tau}$ can be embedded in the minimal extended seesaw framework to allow to relax the cosmological neutrino mass bound $\sum m_\nu \leq 0.12~\text{eV} \mapsto \sum m_\nu \lesssim 1~\text{eV}$.
This is particularly interesting in the light of the sensitivity of upcoming galaxy surveys as \texttt{DESI} and \texttt{EUCLID}.
The 1-$\sigma$ sensitivity of $\sigma\left( \sum m_\nu \right) \simeq 0.02~\text{eV}$ is equivalent to a neutrino mass detection if neutrinos are cosmological stable particles within the cosmological standard model.
A potential non-detection then would be in tension with well established laboratory oscillation experiments, but could be resolved within the neutrino decay scenario.
Furthermore, if the \texttt{KATRIN} experiment reports a neutrino mass detection within its expected sensitivity reach of $\sum m_\nu \simeq 0.6~\text{eV}$, models of neutrino decays are a possible explenation to reconcile cosmological and laboratory measurements.

\section*{Acknowledgments}

The work of SS received the support of a fellowship from ”la Caixa” Foundation (ID 100010434) with fellowship code LCF/BQ/DI19/11730034.

\end{document}